\documentclass[amsmath,amssymb,aps]{revtex4} 
\usepackage{graphicx}
\usepackage{epstopdf}
\usepackage{amsmath}
\usepackage{bm}

\begin{document}

\title{Dual Spacetime Symmetry \\Breaking  down to Einstein Gravity}

\author{Kimihide Nishimura}
\email{kimihiden@dune.ocn.ne.jp}
\affiliation{Nihon-Uniform, 1-4-1 Juso-Motoimazato, Yodogawa-ku Osaka 532-0028 Japan}
\date{\today}

\begin{abstract}
Einstein action of gravity is obtained from a gauge theory, if our spacetime was once in two folds with a double Lorentz symmetry. 
After the dual symmetry breaks spontaneously, Lorentz symmetry absorbs gauge symmetry, while the gauge field begins to drive the vierbein and the spin connection.  
This gauge model of gravity has many bosons and fermions with negative norms, which will be undetectable. 
The consistency of these particles with the Copenhagen interpretation of wave functions, and their relation to the presence of dark matter as well as the absence of antimatter in the universe are discussed. 
\end{abstract}

\maketitle
Comparing Einstein theory of gravity with gauge theories in elementary particle physics, we notice, besides that gravity is extremely weak in microscopic scales, two significant differences; 
one is that gravity has a coupling constant of negative mass dimension, which makes the quantized gravity un-renormalizable, and the other is that the Einstein action of gravity is linear in curvature, while that of gauge field is quadratic in field strength.

Despite these quite incompatible differences, a recent progress in the study of spontaneous spacetime symmetry breakdown shows that a graviton emerges from an SU(2) gauge theory\cite{KN3}, 
although whether the emergent graviton is equivalent to that obtained from Einstein gravity was unclear. 

The emergence of gravity from a gauge theory may become less surprising, if we regard an SU(2) transformation as a local rotation in the isospin space of three dimensions.  
In this respect, we can imagine that an extension of the isospin space to four dimensions  with the Minkowski metric would enable us to obtain Einstein gravity. 
Actually, we show in the following that the loop corrections to an extended isospin gauge theory reproduces, after spacetime symmetry breakdown, the Einstein action of gravity in its complete form as the dominant contribution.

When we extend the isospin space to four dimensions, the number of generators of isospin group does not increase by one, but at least by seven, six of which are identifiable with the generators of the Lorentz transformation in the isospin spacetime. The fundamental representation of the generators of isospin group $I^a=\rho^a/2$, $(a=1,2,3)$ satisfy, besides the algebra of angular momentum: $[I^a, I^b]=i\epsilon_{abc}I^c$, the Clifford algebra:
\begin{equation}
\{I^a, I^b\}=\delta^{ab}/2. 
\end{equation}
We may regard the index of the isospin generators as the coordinate index of the isospin space, and introduce the time component $I^0$ to make it spacetime of four dimensions with the Minkowski metric. 
Then, the four isospin generators $I^\alpha$ will satisfy the algebra:
\begin{equation}
\{I^\alpha, I^\beta\}=-\eta^{\alpha\beta}/2,
\label{ExtendedIsospinAlgebra}
\end{equation}
where the Minkowski metric $\eta^{\alpha\beta}=\eta_{\alpha\beta}$ has signature 
$(+---)$. 
Transforming a superscript to a subscript, or vice versa, of an isospin index is performed by contracting it with $\eta_{\alpha\beta}$ or $\eta^{\alpha\beta}$. 
Due to the resemblance of the relation (\ref{ExtendedIsospinAlgebra}) to that of the Dirac matrices: $\{\gamma^\mu, \gamma^\nu\}=2\eta^{\mu\nu}$, we express $I^\alpha$ by 
\begin{equation}
I^\alpha=\Gamma^\alpha\Gamma_5/2,
\end{equation}
where 
\begin{equation}
\Gamma^\alpha=\left(
\begin{array}{cc}
0&   \bar{\rho}^\alpha  \\
 \rho^\alpha  &0   
\end{array}
\right),
\quad
\Gamma_5=\left(
\begin{array}{lr}
1&   0   \\
0  &-1   
\end{array}
\right),
\quad \rho^\alpha=(1,\rho^a), \quad \bar{\rho}^\alpha=(1,-\rho^a).
\label{IsoDiracMatrices}
\end{equation} 
The matrices $(\Gamma^\alpha, \Gamma_5)$ satisfy $\{\Gamma^\alpha, \Gamma^\beta\}=2\eta^{\alpha\beta}$, $\{\Gamma^\alpha, \Gamma_5\}=0$, and $\{\Gamma_5, \Gamma_5\}=2$. 
The Dirac matrices $(\gamma^\mu,\gamma_5)$ in the chiral representation are expressible in terms of (\ref{IsoDiracMatrices}) by replacing 
$\rho^a$ with the Pauli matrices $\sigma^i$.
Though $\gamma^\mu$ and $\Gamma^\alpha$ are expressed commonly by $4\times4$ matrices, they operate on different spinor spaces, and therefore commute: $[\gamma^\mu, \Gamma^\alpha]=0$. 

The algebra of $I^\alpha$ does not close, and the commutators $[I^\alpha, I^\beta]$ represent new generators. If we define the additional six generators by $I^{\alpha\beta}=[\Gamma^\alpha, \Gamma^\beta]/4$, 
the algebra closes as follows:
\begin{equation}
[I^\alpha, I^\beta]=-I^{\alpha\beta},\quad
[I^\alpha,I^{\beta\gamma}]=\eta^{\alpha\beta}I^\gamma-\eta^{\alpha\gamma}I^\beta,
\quad
[I^{\alpha\beta},I^{\gamma\delta}]=
\eta^{\alpha\delta}I^{\beta\gamma}+\eta^{\beta\gamma}I^{\alpha\delta}
-\eta^{\alpha\gamma}I^{\beta\delta}-\eta^{\beta\delta}I^{\alpha\gamma}.
\end{equation}
The 10 generators $(I^\alpha, I^{\alpha\beta})$ are not Hermitian, but ``quasi Hermitian"  or ``quasi anti-Hermitian" in the following sense:
\begin{equation}
\bar{I}^\alpha:=\Gamma^0(I^\alpha)^\dagger\Gamma^0=I^\alpha, \quad
\bar{I}^{\alpha\beta}:=\Gamma^0(I^{\alpha\beta})^\dagger\Gamma^0=-I^{\alpha\beta}.
\end{equation}
A gauge theory invariant under the 10 dimensional isospin group is constructible in terms of a quartet of Dirac spinors: 
$\psi=(\psi_1,\psi_2)^T=(\psi_{11},\psi_{12},\psi_{21},\psi_{22})^T$ 
transforming as 
\begin{equation}
\delta\psi=i\theta\psi,\quad \delta\bar{\psi}=-i\bar{\psi}\theta, \quad 
\bar{\psi}:=\psi^\dagger\gamma^0\Gamma^0, \quad 
\theta:=\theta_\alpha I^\alpha-\frac{i}{2}\theta_{\alpha\beta}I^{\alpha\beta},
\end{equation}
where $(\theta_\alpha,\theta_{\alpha\beta})$ are infinitesimal 10 real gauge parameters.
The Lagrangian for the Dirac quartet is given by
\begin{equation}
{\cal L}_F=\bar{\psi}\gamma^\mu iD_\mu\psi, \quad 
D_\mu:=\partial_\mu+igY_\mu, \quad
Y_\mu:=Y_{\mu\alpha}I^\alpha-\frac{i}{2}Y_{\mu\alpha\beta}I^{\alpha\beta}, 
\label{L_F}
\end{equation}
where the coupling constant $g$ and the Yang-Mills gauge fields $(Y_{\mu\alpha}, Y_{\mu\alpha\beta})$ are real. 
The Lagrangian (\ref{L_F}) is gauge invariant, 
if $Y_\mu$ transforms as $\delta Y_\mu=-g^{-1}\partial_\mu\theta+i[\theta, Y_\mu]$, 
or explicitly,
\begin{equation}
\begin{array}{lcl}
\delta Y_{\mu\alpha}&=&-g^{-1}\partial_\mu\theta_\alpha
-Y_{\mu\alpha\beta}\theta^\beta+\theta_\alpha{}^\beta Y_{\mu\beta},\\
\delta Y_{\mu\alpha\beta}&=&-g^{-1}\partial_\mu\theta_{\alpha\beta}
+\theta_\alpha{}^\gamma Y_{\mu\gamma\beta}
+\theta_\beta{}^\gamma Y_{\mu\alpha\gamma}
+\theta_\alpha Y_{\mu\beta}-\theta_\beta Y_{\mu\alpha}. 
\end{array}
\label{GTY}
\end{equation}
The field strength $F_{\mu\nu}$ are obtained from the commutator of the covariant derivatives: $[D_\mu,D_\nu]=igF_{\mu\nu}$: 
\begin{equation}
\begin{array}{rcl}
F_{\mu\nu}&:=&F_{\mu\nu\alpha}I^\alpha-\frac{i}{2}F_{\mu\nu\alpha\beta}I^{\alpha\beta},\\
F_{\mu\nu\alpha}&=&\partial_\mu Y_{\nu\alpha}-\partial_\nu Y_{\mu\alpha}
+gY_{\mu\alpha}{}^\beta Y_{\nu\beta}-gY_{\nu\alpha}{}^\beta Y_{\mu\beta},\\
F_{\mu\nu\alpha\beta}&=&
\partial_\mu Y_{\nu\alpha\beta}-\partial_\nu Y_{\mu\alpha\beta}
+gY_{\mu\alpha}{}^\gamma Y_{\nu\gamma}{}_\beta
-gY_{\nu\alpha}{}^\gamma Y_{\mu\gamma}{}_\beta
+gY_{\mu\alpha}Y_{\nu\beta}-gY_{\nu\alpha}Y_{\mu\beta}.
\end{array}
\label{FieldStrength}
\end{equation}
Since the gauge transformation of $F_{\mu\nu}$ is given by
$\delta F_{\mu\nu}=[i\theta, F_{\mu\nu}]$: 
\begin{equation}
\begin{array}{lcl}
\delta F_{\mu\nu\alpha}&=&-F_{\mu\nu\alpha\beta}\theta^\beta
+\theta_\alpha{}^\beta F_{\mu\nu\beta},\\
\delta F_{\mu\nu\alpha\beta}&=&
\theta_\alpha F_{\mu\nu\beta}-\theta_\beta F_{\mu\nu\alpha}
+\theta_\alpha{}^\gamma F_{\mu\nu\gamma\beta}
+\theta_\beta{}^\gamma F_{\mu\nu\alpha\gamma},
\label{GTF}
\end{array}
\end{equation}
the following quadratic form is gauge invariant:
\begin{equation}
-\frac{1}{4}{\rm Tr}(F^{\mu\nu}F_{\mu\nu})=
\frac{1}{4}F^{\mu\nu\alpha}F_{\mu\nu\alpha}
-\frac{1}{8}F^{\mu\nu\alpha\beta}F_{\mu\nu\alpha\beta}.
\end{equation}
We reconsider here the kinetic term for the gauge fields. If our theory is re-normalizable, 
the divergent part of renormalization by fermion loops should coincide in form with the original Lagrangian for the gauge fields.  
In this sense, it is not necessary for us to give by hand the Lagrangian for the gauge fields; it will be obtained from the loop corrections by the Dirac quartet. 
In fact, if finite terms are omitted, the path integral formulation of quantum field theory gives 
for the action of gauge fields:
\begin{equation}
\begin{array}{ll}
&-i{\rm Tr}\ln[1-(g\gamma\cdot Y)(i\gamma\cdot\partial)^{-1}]\\
=&\displaystyle\int d^4x\left[
g^2K_1Y^{\mu\alpha}Y_{\mu\alpha}
-(g^2K_1/2)Y^{\mu\alpha\beta}Y_{\mu\alpha\beta}
-(4g^2K_2/3)\left(\frac{1}{4}F^{\mu\nu\alpha}F_{\mu\nu\alpha}
-\frac{1}{8}F^{\mu\nu\alpha\beta}F_{\mu\nu\alpha\beta}\right)+\cdots\right], 
\end{array}
\label{LoopCorrecrion}
\end{equation}
where the constants $K_1$ and $K_2$ are quadratically and  logarithmically divergent, respectively, which are symbolically given by
\begin{equation}
K_n:=\int\frac{d^4p}{(2\pi)^4}\frac{i}{(p^2+i\epsilon)^n}.
\end{equation}
The estimation of divergent integrals using some appropriate regulators usually gives a positive constant or zero for $K_1$, and a negative constant for $K_2$. 
The term proportional to $K_2$ is gauge invariant, while those proportional to $K_1$ are symmetry breaking. 

The assignment of constant values to two $K_1$s  in (\ref{LoopCorrecrion}) determines the pattern of symmetry breaking. 
Since we are here interested in the symmetry breaking of the type which preserves Lorentz symmetry in the isospin spacetime, the Lagrangian of the gauge field is given by 
\begin{equation}
{\cal L}_G=\frac{3}{2}g^2\phi_0^2Y^{\mu\alpha}Y_{\mu\alpha}
+\frac{1}{4}F^{\mu\nu\alpha}F_{\mu\nu\alpha}
-\frac{1}{8}F^{\mu\nu\alpha\beta}F_{\mu\nu\alpha\beta},
\label{L_G}
\end{equation}
where $\phi_0$ is a constant of mass dimension one. 

Up to now, we are formulating the model in a flat Minkowski spacetime. 
In order to consider spontaneous spacetime symmetry breakdown preserving Lorentz invariance, 
it is necessary for us to reformulate the model in a general coordinate system by introducing the vierbein $e_\mu{}^\alpha$ and the local Lorentz connection $\omega_{\mu\alpha\beta}$ as auxiliary fields. 
The requirement that the vacuum expectation value $\langle Y_{\mu\alpha}\rangle$ should not violate the general covariance and the local Lorentz covariance will be satisfied by postulating that  
\begin{equation}
\langle Y_{\alpha\beta}\rangle=\langle e^\mu{}_\alpha Y_{\mu\beta}\rangle=\phi_0\eta_{\alpha\beta}.
\label{InterlockCondition}
\end{equation}
Despite using the same greek letters, the first index of $Y_{\alpha\beta}$ is the local Lorentz index, while the second is the isospin index. After symmetry breaking, these two indices are indistinguishable, which requires additional consistency conditions. 
First, the Lorentz invariance of the condition (\ref{InterlockCondition}) requires that the local Lorentz transformation 
$\delta e^\mu{}_\alpha=\epsilon_\alpha{}^\beta e^\mu{}_\beta$ should be accompanied by the gauge transformation $\theta_{\alpha\beta}=\epsilon_{\alpha\beta}$ . 
Second, in order that the gauge transformation by $\theta_{\alpha\beta}$ in (\ref{GTY}) and (\ref{GTF}) are identifiable  with the local Lorentz transformation, $gY_{\mu\alpha\beta}$ should equal to the local Lorentz connection $\omega_{\mu\alpha\beta}$. 
Accordingly, spontaneous spacetime symmetry breakdown preserving the general covariance and the local Lorentz covariance realizes for 
\begin{equation}
Y_{\mu\alpha}=\phi e_{\mu\alpha}, \quad Y_{\mu\alpha\beta}=g^{-1}\omega_{\mu\alpha\beta},
\label{MatchingConditions}
\end{equation}
where $\phi$ is some scalar field of mass dimension one 
with the vacuum expectation value $\langle\phi\rangle=\phi_0$.  
These relations dictate how the gauge fields drive gravity after symmetry breaking.  

Due to the relations (\ref{MatchingConditions}), we find that 
\begin{equation}
F_{\mu\nu\alpha}=(\partial_\mu\phi)e_{\nu\alpha}-(\partial_\nu\phi)e_{\mu\alpha}
+\phi(\nabla_\mu e_{\nu\alpha}-\nabla_\nu e_{\mu\alpha}),
\label{Fmunualpha}
\end{equation}
where the covariant derivative of the vierbein is defined by
\begin{equation}
\nabla_\mu e_{\nu\alpha}:=\partial_\mu e_{\nu\alpha}
-\Gamma^\rho_{\mu\nu}e_{\rho\alpha}+\omega_{\mu\alpha}{}^\beta e_{\nu\beta},
\end{equation}
with the Christoffel symbol $\Gamma^\rho_{\mu\nu}$ calculated for the metric $g_{\mu\nu}=e_\mu{}^\alpha e_{\nu\alpha}$.
The last term proportional to $\phi$ in (\ref{Fmunualpha}) will vanish, if the vierbein satisfies the torsion constraint, which implies that the second term in (\ref{L_G}) resolves the constraint dynamically.  
If the torsion constraint is imposed, we have $\nabla_\mu e_{\nu\alpha}=0$, and 
\begin{equation}
F_{\mu\nu\alpha\beta}=g^{-1}R_{\mu\nu\rho\sigma}e^\rho{}_\alpha e^\sigma{}_\beta
+g\phi^2(e_{\mu\alpha}e_{\nu\beta}-e_{\nu\alpha}e_{\mu\beta}), 
\end{equation}
where $R_{\mu\nu\rho\sigma}$ is the Riemann curvature.   
As the result, the Lagrangian of the gauge field reduces to
\begin{equation}
{\cal L}_G=\frac{3}{2}g^{\mu\nu}\partial_\mu\phi\partial_\nu\phi
-3g^2(\phi^2-\phi_0^2)^2-\frac{\phi^2}{2}R
-\frac{1}{8g^2}R^{\mu\nu\rho\sigma}R_{\mu\nu\rho\sigma}, 
\label{LagOfGravity}
\end{equation}
where $R=e^{\mu\alpha}e^{\nu\beta}R_{\mu\nu\alpha\beta}$ is the scalar curvature. 
We have added a constant term $-3g^2\phi_0^4$ in (\ref{LagOfGravity}) for convenience. 
The potential term of $\phi$ attains its minimum for $\phi=\phi_0$. 
Then, redefining the scalar field by $\phi=\phi_0+\sigma/\sqrt{3}$ gives
\begin{equation}
{\cal L}_G=\frac{1}{2}g^{\mu\nu}\partial_\mu\sigma\partial_\nu\sigma
-\frac{m_\sigma^2}{2}\sigma^2
-g\sqrt{\frac{2}{3}}m_\sigma\sigma^3
-\frac{g^2\sigma^4}{3}
-\frac{R}{16\pi G}\left(1+\sqrt{\frac{8}{3}}\frac{g\sigma}{m_\sigma}\right)^2
-\frac{1}{8g^2}R^{\mu\nu\rho\sigma}R_{\mu\nu\rho\sigma},
\end{equation}
where 
\begin{equation}
G=\frac{\alpha_G}{m_G^2}, \quad \alpha_G=\frac{g^2}{4\pi}, \quad m_G=\frac{m_\sigma}{2}=\sqrt{2}g\phi_0.
\end{equation}

The $R$-term, representing the Einstein gravity, will dominate comparing with the quadratic term: $R^{\mu\nu\rho\sigma}R_{\mu\nu\rho\sigma}$, if $m_G$ is large enough. 
Then, $G$ should equal to the Newton constant.  
Thus, the Einstein action of gravity emerges from a gauge theory of a Dirac quartet. 

The additional massive scalar field $\sigma$ emergent in association with gravity 
plays a role similar to the Higgs boson. 
In fact, after symmetry breaking, the Dirac quartet (\ref{L_F}) has the Lagrangian: 
\begin{equation}
{\cal L}_F=\bar{\psi}\left[e^\mu{}_\alpha\gamma^\alpha i\nabla_\mu-M\right]\psi
 -\sqrt{\frac{2}{3}}\frac{g\sigma}{m_G}\bar{\psi}M\psi,
\quad
\nabla_\mu=\partial_\mu
+\frac{1}{2}\omega_{\mu\alpha\beta}(
\Sigma^{\alpha\beta}+I^{\alpha\beta}),
\quad
M:=\frac{m_G}{\sqrt{2}}\gamma^\alpha I_\alpha,
\label{QDL}
\end{equation}
where 
$\Sigma^{\alpha\beta}=[\gamma^\alpha,\gamma^\beta]/4$. 
We notice that the ordinary spin connection  
$\omega_{\mu\alpha\beta}\Sigma^{\alpha\beta}/2$ has appeared in the covariant derivative. The $\sigma$-boson couples to the mass term of the Dirac quartet as the Higgs boson couples to the mass terms of leptons and quarks in the standard theory of elementary particles.  
We may call in this respect the scalar boson $\sigma$ the gravitational Higgs boson, or the second Higgs boson.

Incidentally, in our formulation of the isospin gauge theory, the Minkowski spacetime and the isospin spacetime is interchangeable. 
It is easy to verify that the operator $T$ defined by $T:=\bar{\rho}^\alpha\sigma_\alpha/2=\rho^\alpha\bar{\sigma}_\alpha/2$ satisfies 
\begin{equation}
T^2=1,\quad T\gamma^\alpha T=\Gamma^\alpha, \quad T\Gamma^\alpha T=\gamma^\alpha.
\end{equation}
which transforms spins into isospins, and vice versa, and enables us to represent the Lagrangian of the Dirac quartet 
in the isospin spacetime. 
Defining $\psi':=T\psi$, and $\bar{\psi}':=\bar{\psi}T$, 
we have from (\ref{L_F}) that  
\begin{equation}
{\cal L}_F=\bar{\psi}'\Gamma^\mu iD'_\mu\psi',\quad 
D'_\mu:=\partial_\mu+igY_{\mu\alpha}\Sigma^\alpha+\frac{g}{2}Y_{\mu\alpha\beta}\Sigma^{\alpha\beta}, 
\label{DDL}
\end{equation}
since $TI^\alpha T=\Sigma^\alpha:=\gamma^\alpha\gamma_5/2$, and 
$TI^{\alpha\beta}T=\Sigma^{\alpha\beta}$. 

Before the duality transformation, Lorentz invariance was spacetime symmetry associated with the change of coordinates $\delta x^\mu=\epsilon^\mu{}_\nu x^\nu$, while iso-Lorentz invariance was gauge symmetry without coordinate changes. 
The duality operator $T$ interchanges these two symmetries, where iso-Lorentz invariance of (\ref{DDL}) is accompanied by the coordinate change $\delta x^\mu=\theta^\mu{}_\nu x^\nu$. 
Lorentz invariance becomes in turn gauge symmetry without coordinate changes.
Therefore in the dual representation, $x^\mu$ is interpreted as the iso-spacetime coordinate, while $\alpha$ and $\beta$ are gauge indices of an extended Lorentz generators $(\Sigma^\alpha,\Sigma^{\alpha\beta})$.
The duality suggests that the isospin spacetime also may have particles and energies of its own, and constitute another universe. 

Whereas we do not reconsider from the viewpoint of our gauge theory the difficulties known in quantizing the Einstein gravity\cite{DeWitt,Ashtekar,Rovelli},  
we consider instead the problem arising from quantizing a gauge theory of a non-compact gauge group, where the negative norms and the negative probabilities arise.

Quantizing the gauge field in the ghost-less gauge\cite{Feynman}: $Y_{0\alpha}=Y_{0\alpha\beta}=0$,
we find from (\ref{L_G}) that the gauge field components
$Y_{i0}$ and $Y_{ia0}$ have negative kinetic terms, 
and therefore the corresponding bosons have negative norms. 
A similar distinction appears also for the Dirac quartet.   
Rewriting the kinetic part of (\ref{L_F}) by
\begin{equation}
\bar{\psi}i\gamma\cdot\partial\psi=\bar{\psi}_+i\gamma\cdot\partial\psi_+-\bar{\psi}_-i\gamma\cdot\partial\psi_-, \quad \psi_\pm:=(\psi_1\pm\psi_2)/\sqrt{2}, \quad \bar{\psi}_\pm:=\psi_\pm^\dagger\gamma^0, 
\label{psi_pm}
\end{equation}
we find that the Dirac doublet $\psi_-$ has the negative kinetic term, and the corresponding fermions have negative norms. 

The bosons and fermions with negative norms still have positive energy eigenvalues and positive energy expectation values as ever. 
Further, the completeness of eigenstates of a Hermitian operator, as well as the unity of the total probability of transitions, or of time evolution, still hold in the following form:
\begin{equation}
\sum_iP_i=1, \quad P_i=\frac{\vert \langle i\vert\Psi\rangle\vert^2}{\langle i\vert i\rangle \langle\Psi\vert \Psi\rangle},\quad 
1=\sum_i\frac{\vert i\rangle\langle i\vert}{\langle i\vert i\rangle}, \quad
H\vert i\rangle=E_i\vert i\rangle, 
\end{equation}
where the complete set $\vert i\rangle$ and the wave function $\vert\Psi\rangle$ are not normalized. 

A problem arises when we consider the transition of a positive norm state to, or from, a negative norm state, 
where $P_i\leq0$ due to $\langle i\vert  i\rangle \langle\Psi\vert \Psi\rangle<0$ for some $i$, since negative probabilities will contradict the Copenhagen interpretation of wave functions. However, there are two orthogonal sets of physical states 
$\{\vert\Psi_+\rangle\}$ and $\{\vert\Psi_-\rangle\}$, satisfying 
$\langle\Psi_+\vert \Psi_+\rangle>0$, $\langle\Psi_-\vert \Psi_-\rangle<0$, and 
$\langle\Psi_+\vert \Psi_-\rangle=0$, which do not conflict with the probability interpretation. 

Denoting quantum fields with positive norm by $Y_+$ for bosons, and $\psi_+$ for fermions, while those with negative norms by $Y_-$, and $\psi_-$,
we see that every interaction term contained in ${\cal L}_F+{\cal L}_G$ has the negative norm fields in even number or zero:
$Y_+\psi_+\psi_+$, $Y_+\psi_-\psi_-$, $Y_-\psi_+\psi_-$, $Y_+Y_+Y_+$, $Y_+Y_-Y_-$, 
$Y_+Y_+Y_+Y_+$, $Y_+Y_+Y_-Y_-$, and $Y_-Y_-Y_-Y_-$. 
This implies that the signature of the norm of any particle number eigenstate does not change by time evolution.
Then, we have $\langle i_+\vert\Psi_-\rangle=\langle i_-\vert\Psi_+\rangle=0$, 
where the wave function $\vert\Psi_+\rangle$ is spanned only by the particle number eigenstates with positive norms, 
while $\vert\Psi_-\rangle$ is only by those with negative norms. The positive norm wave function $\vert\Psi_+\rangle$ may contain even number of negative norm particles, while the negative norm wave function $\vert\Psi_-\rangle$ may contain, besides odd number of negative norm particles, arbitrary number of positive norm particles. 

Consequently, the Hilbert space of the state vectors consistent with the Copenhagen interpretation is divided into two distinct sectors: ${\rm H}={\rm H}_+\cup{\rm H}_-$, and the wave function of the universe will generally be the direct product of the positive and negative norm wave functions: 
$\vert\Psi\rangle=\vert\Psi_+\rangle\vert\Psi_-\rangle$.
In view of the duality of our spacetime and the isospin spacetime, it is intuitive to assign $\vert\Psi_-\rangle$ to the isospin spacetime.
Then, the orthogonality relations $\langle i_+\vert\Psi_-\rangle=0$, and $\langle i_-\vert\Psi_+\rangle=0$ will be interpreted as that the particles and energies in the isospin spacetime are undetectable by gauge interactions from the observer in our spacetime, and vice versa. 

The bosons with negative norms seem to exist only before the symmetry breaking. 
After symmetry breaking, the gauge bosons turn into the massive scalar boson and the gravitons, all of which have positive norms. One interpretation of this observation is that all the gauge bosons with negative norms were unphysical, and eliminated by the torsion constraint and some gauge fixings. 
Another interpretation is that the negative norm particles were transferred into the negative norm sector of the universe $\vert\Psi_-\rangle$, and have become not detectable.

Concerning the fermions, on the other hand, if the Lagrangian (\ref{QDL}) is decomposed into a left-handed Weyl quartet and a right-handed Weyl quartet,  
we obtain the particle spectrum:  
\begin{equation}
\begin{array}{cc}
p^0_{L+}=\left(p+\delta, -(p-\delta), p, -\omega_G\right), &
p^0_{R+}=\left(p-\delta, -(p+\delta), -p, \omega_G\right), \\
p^0_{L-}=\left(p-\delta, -(p+\delta), -p, \omega_G\right), &
p^0_{R-}=\left(p+\delta, -(p-\delta), p, -\omega_G\right), 
\end{array}
\label{PS}
\end{equation}
where $\omega_G=\sqrt{p^2+m_G^2}$, and $p^0_{L+}$ and $p^0_{L-}$ represent the energies for the left-handed Weyl quartet with momentum $p$. The signature ``+" implies the positive norm states, while ``$-$" the negative norm states. 
Similarly, $p^0_{R+}$ and $p^0_{R-}$ are for the right-handed Weyl quartet. 
An infinitesimal chemical potential $\delta>0$ has been introduced by hand to the massless zero norm fermions in order to recategorize these particles into positive or negative norm particles. The zero norm fermions will be actually unphysical, since the expectation value of the energy is zero.
The negative energies in (\ref{PS}) represent the antiparticles with energies with the opposite sign. 
As the result, a Dirac quartet turns into two unphysical, one massless, and one massive Dirac fermions after symmetry breaking. 
Another feature drawn from this particle spectrum is that a particle and its antiparticle possess opposite norms in each chiral multiplet.
This observation seems to have a relation to the problem of antimatter almost absent in the universe\cite{Sakhalov}, 
since then the universe consisting of the positive norm particles can contain only particles or only antiparticles for each kind. 
The situation will be realized when almost all the antiparticles belong to the negative norm sector of the universe $\vert\Psi_-\rangle$. Then, antiparticles are rarely detected from the observer in the positive norm sector of the universe $\vert\Psi_+\rangle$.

We can further examine this issue from another point of view. 
The emergence of negative norm states or zero norm states in quantum mechanics will indicate that there are redundancies in describing the system.
Then, imposing some constraint will eliminate those undesired states. 
This is in fact achieved in our case by imposing the constraint $\psi_-=0$ for the Dirac quartet.
Then, the Lagrangian (\ref{L_F}) reduces to an SU(2)$\times$SU(2) model of a Dirac doublet:
\begin{equation}
{\cal L}_F^{(+)}=\bar{\psi}_+\gamma^\mu\left(i\partial_\mu+\frac{g\rho^a}{2}(X_{\mu a}-Y_{\mu a})
\right)\psi_+, \quad X_{\mu a}:=\frac{1}{2}\epsilon_{abc}Y_{\mu bc}, 
\end{equation}

where $X_{\mu a}$ are massless SU(2) vector bosons, while $Y_{\mu a}$ are massive. 
The particle spectrum is then,
\begin{equation}
p^0_{L+}=\displaystyle\left(p+\frac{m}{2}, -(p-\frac{m}{2}), \omega-\frac{m}{2},-(\omega+\frac{m}{2})\right), 
p^0_{R+}=\displaystyle\left(p-\frac{m}{2}, -(p+\frac{m}{2}), \omega+\frac{m}{2},-(\omega-\frac{m}{2})\right), 
\end{equation}
where $\omega=\sqrt{p^2+m^2}$ and $m=g\phi_0$.
The result returns to that obtained in the previous papers\cite{KN3,KN1,KN2}. 
Whereas all the fermions and gauge bosons have positive norms, the chemical potential of a fermion is the opposite to that of its anti-fermion. Then, also in this case, either the particles or the antiparticles will dominate under thermal equilibrium, if the quasi fermion number violating processes are allowed in course of symmetry breaking. 

Even in this reduced model, gravitons appear after symmetry breaking, but Einstein gravity is incompletely reproduced\cite{KN3}. 
In this respect, the existence of bosons and fermions with negative norms will be unavoidable, if Einstein gravity is to be founded on quantum mechanics.  

The negative norm sector of the universe $\vert\Psi_-\rangle$, which has no gauge interactions with the positive norm universe $\vert\Psi_+\rangle$, will interact with $\vert\Psi_+\rangle$ through gravity after symmetry breaking,   
and therefore the classical equation of gravity will have the form:
\begin{equation}
R^{\mu\nu}-\frac{1}{2}g^{\mu\nu}R=8\pi G\left(T^{\mu\nu}_++T^{\mu\nu}_-\right), \quad
T^{\mu\nu}_\pm:=\frac{\langle\Psi_\pm\vert T^{\mu\nu}\vert\Psi_\pm\rangle}{\langle\Psi_\pm\vert\Psi_\pm\rangle},
\end{equation}
where $\vert\Psi_-\rangle$ is to describe dark side of the universe responsible for the dark matter and the dark energy in cosmology\cite{Weinberg1,Weinberg2,Hinshaw}.  
Though quantum gravity may threaten again the probability interpretation of state vectors, it will be suppressed by the Planck mass scale, if exists.

Our gauge model of gravity predicts at least three new particles detectable in principle; 
the gravitational Higgs boson $H_G$ with mass $2m_G$, and the gravitational Dirac doublet, the ``gravino" and the ``gravion": ($\nu_{\rm G}$, G) with mass zero and $m_G$, which may also be the dark matter. 

\end{document}